\documentstyle[festkoer,psfig]{vieweg}

\author{Bodo Huckestein${}^{1,2}$, Michael Backhaus${}^2$}

\Kurzautor{Huckestein et al.}

\title{Scaling in the Integer Quantum Hall Effect: interactions and
  low magnetic fields}

\Kurztitel{Scaling in the IQHE}

\Adresse{${}^1$ Theoretische Physik III, Ruhr-Universit\"at Bochum,
  Universit\"atsstra\ss{}e 150, D-44780 Bochum, Germany\\
  ${}^2$ Institut f\"ur Theoretische Physik, Universit\"at zu K\"oln,
  Z\"ulpicher Str. 77, D-50937 K\"oln, Germany}

\begin{document}

\Titel

\begin{abstract}
  Recent developments in the scaling theory of the integer quantum
  Hall effect are discussed. In particular, the influence of
  electron-electron interactions on the critical behavior are
  studied. It is further argued that recent experiments on the
  disappearance of the quantum Hall effect at low magnetic fields support
  rather than disprove the scaling theory, when interpreted properly.
\end{abstract}

While the remarkable accuracy of the quantization of the Hall
conductivity is the most prominent feature of the integer quantum Hall 
effect (IQHE) \cite{KDP80}, the nature of the transitions between different
quantized plateaus is one of the most interesting aspects of the
effect \cite{Huc95r}. The formation of the plateaus is due to the
localization of the charge carriers by a residual disorder potential
and the transition between the plateaus corresponds to a
delocalization transition near the centers of the Landau bands. 
Despite two decades of research the correct field theoretical
description of localization the IQHE is still controversial and to a
large extent speculative \cite{Zir99}. Even the essential ingredients
of such a theory are not evident. In a real sample, the electrons are
interacting, their wavefunction have a finite extent perpendicular to
the plane of electron gas, and they move in a finite magnetic
field. In contrast, the simplest theories assume strictly
two-dimensional, non-interacting electrons moving in a magnetic field
that is sufficiently strong to neglect Landau level mixing. In the
present contribution we will discuss to what extent the Coulomb
interactions between the electrons and strong Landau level mixing
change the critical behavior associated with the plateau transitions.

The origin of the quantization of the Hall conductivity is an
excitation gap in the clean system. In the IQHE this gap is the gap
between different Landau levels, and hence of single particle origin,
while in the fractional QHE electron correlations lead to a gap at
magical filling factors \cite{Lau83}. Since electron-electron
interactions are unimportant for the origin of the quantization it is
tempting to try to describe the IQHE by a theory of non-interacting
electrons. Furthermore, since the quantization is most prominent at
high magnetic fields where the cyclotron energy is much larger than
the disorder, a projection onto a single Landau level and the neglect
of Landau level mixing seems reasonable. Unfortunately, even such a
simplified model still resist analytic approaches. However, it is
quite amenable to numerical simulations. The results compare quite
favorably to experiments. In particular, they show evidence for a
critical point near the center of each disorder broadened Landau band.
Near the critical point single parameter scaling holds with a single
relevant scaling index, the localization length exponent $\nu\approx
2.35\pm0.03$ \cite{HK90,Huc95r}. In fact, this value agrees with the
experimental value of $\nu\approx 2.3\pm0.1$ \cite{KHKP91b}.
Simulations show that the dissipative conductivity at the critical
point in the lowest Landau band takes on a universal value of
$\sigma_{xx}\approx 0.5 e^2/h$ \cite{HHB93}, in agreement with some
experiments \cite{Sea95}.

Despite these successes, a serious discrepancy exists between the
theory for non-interacting electrons and experiments regarding the
value of the dynamical critical exponent $z$. For non-interacting
electrons $z$ equals to the space dimension $d=2$ and the width of the
transition region is then expected to scale as a power of an
external frequency $\omega$ or temperature $T$ with an exponent
$1/z\nu\approx 0.21$. In contrast, experiments show power law
behavior, but with an exponent of about 0.42 \cite{WTPP88,ESKT93},
compatible with a dynamical critical exponent of $z\approx1$.
Apparently, the non-interacting theory is an inadequate description of
the real world and in Section \ref{sec:Electr-electr-inter} we
will discuss how the critical properties of the theory change upon
inclusion of the effects of electron-electron interactions.

In addition to the effects of electron-electron interactions one has
to consider the effects of a finite strength of the applied magnetic
field and in particular the fate of the QHE as the magnetic field is
switched off. Field-theoretic arguments suggest that the isolated
critical points persist in the presence of Landau level coupling
\cite{LLP83}, though not necessarily anymore at the centers of the
Landau levels. However, for vanishing magnetic field the scaling
theory of Anderson localization tells us that all states below the
Fermi energy are localized in two dimensions \cite{AALR79}. To
reconcile this two results, Khmelnitskii and Laughlin have argued that
the critical energies float upwards in energy as the magnetic field
turns to zero and eventually move through the Fermi energy so that all
states below are localized \cite{Khm84,Lau84}. In this levitation
scenario the critical states move consecutively through the Fermi energy
and the last transition to the insulator always happens from the
plateau with Hall conductivity $ne^2/h$ with $n=1$. Recently, this
picture has been questioned by experiments that seem to show direct
transitions from higher plateaus, $n=2,\ldots,6$, to the insulator
\cite{SKD93,KMFP95,Sea97,LCSL98,Hea99}. As experiments are performed
at finite temperatures and hence probe finite length scales and
scaling theory deals with diverging length scales, it is important in
this context to consider the implications of scaling for finite
systems and temperatures. This will be done in Section
\ref{sec:Fate-QHE-at} \cite{Huc00l}.

\section{Electron-electron interactions}
\label{sec:Electr-electr-inter}

\subsection{Relevance of interactions}
\label{sec:Relev-inter}

We will use the language of the renormalization group for our
discussion of the influence of interactions. In the absence of
interactions, the plateau transitions correspond to quantum critical
points that are characterized by a set of universal exponents, the
localization length exponent $\nu$, the dynamical critical exponent
$z$, the spectrum of generalized multifractal dimensions $D(q)$, where 
the exponent $D(2)=2-\eta$ enters the dynamical density correlator
\cite{CD88}, as well as the critical conductivity $\sigma_c$
\cite{HHB93}. Numerical evidence suggests that these critical
quantities are independent of microscopic details like the Landau
level index or the nature of the disorder potential if corrections to
scaling are properly taken into account \cite{Huc93l}.

The results are obtained for models of disordered non-interacting
electrons in two dimensions. The finite extent of the electron
wavefunction in the direction perpendicular to the two-dimensional
plane does not change the scaling behavior as it only modifies the
form factor of the disorder as long as only a single subband of the
perpendicular motion is occupied. We will now consider the effect of
an additional interaction potential
\begin{equation}
  \label{eq:1}
  V(r) = A r^{-\lambda}
\end{equation}
between the electrons. The physical Coulomb interaction corresponds to
$y=1$ and $A=e^2/4\pi\epsilon$. In order to determine the relevance of
this perturbation at the non-interacting fixed point of the
renormalization group, we calculate the disorder average
$\overline{V}$ of the interaction between two consecutive
non-interacting eigenstates and compare this to the mean level spacing
$\Delta$.

\begin{figure}
  \hfill\psfig{figure=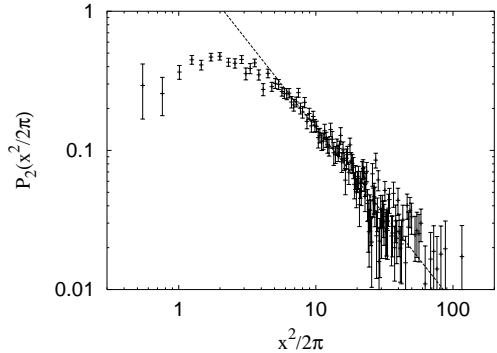,width=7.0cm}\hspace*{\fill}
  \caption[fig1]{The Fourier transform of the correlation function
    $P_2(r,L_\omega)$ as a function of $x=qL_\omega$. System sizes
    containing from 169 to 900 flux quanta were used with up to 100
    realizations of the disorder potential. The dashed line
    corresponds to $\tilde{\eta}=-0.5$.}
  \label{fig:1}
\end{figure}

The average interaction $\overline{V}$ is just the disorder average of
the Hartree-Fock operator,
\begin{equation}
  \label{eq:5}
  \overline{V} = \int d^2r P_2(\mathbf{r},\omega\to0) V(\mathbf{r}),
\end{equation}
\begin{eqnarray}
  \lefteqn{P_2(r,\omega) = }\nonumber\\
  &&\overline{\sum_{i,j}
    \left(|\psi_i(0)|^2|\psi_j(\mathbf{r})|^2 -
      \psi^*_i(0)\psi_i(\mathbf{r})\psi^*_j(\mathbf{r})\psi_j(0)\right)
    \delta(E_i-E_j-\hbar\omega)},  
  \label{eq:6}
\end{eqnarray}
where $E_i<E_c$ and $E_j>E_c$, the sum runs over all eigenstates 
of the non-interacting system, and $E_c$ is the critical energy. At the 
critical point $P_2(r,\omega)$ shows power law scaling,
\begin{equation}
  \label{eq:7}
  P_2(r,\omega) \propto \left(\frac{r}{L_\omega}\right)^{-\tilde{\eta}},
\end{equation}
with $L_\omega=(\rho_0\hbar\omega)^{-1/2}$, the effective system size
with level spacing $\hbar\omega$. The exponent $\tilde{\eta}$ reflects
the multifractal character of the critical eigenstates.
Fig.~\ref{fig:1} shows numerical results for the scaling of the
Fourier transform $P_2(qL_\omega)$ in the limit $q,\omega\to0$. From
the asymptotic power law $P_2(qL_\omega)\propto
(qL_\omega)^{-2+\tilde{\eta}}$, we obtain $\tilde{\eta}=-0.5\pm0.1$.

\begin{figure}
  \hfill\psfig{figure=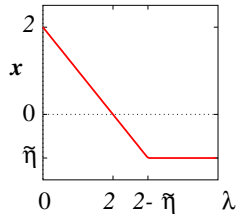,width=3.0cm}\hspace*{\fill}
  \caption[fig2]{The scaling dimension $x$ of the interaction as
    function of its range exponent $\lambda$.}
  \label{fig:2}
\end{figure}

The average interaction $\overline{V}$ scales then like
\begin{equation}
  \label{eq:8}
  \overline{V}\propto A \left(L^{-\lambda} -
    L^{\tilde{\eta}-2}\right).
\end{equation}
This energy scale has to be compared to the mean level spacing of the
non-interacting system $\Delta=(\rho_0L^2)^{-1}$. If $\overline{V}$
scales faster to zero than $\Delta$ with increasing system size then
the non-interacting eigenstates are essentially unaffected by the
interaction and the interaction is irrelevant, while in the opposite
case increasingly more eigenstates get coupled with increasing system
size and the interaction is relevant in the renormalization group
sense. Defining the scaling index $x$ of the interaction by
\begin{equation}
  \label{eq:9}
  \frac{\overline{V}}{\Delta}\propto L^{x},
\end{equation}
we see that for $\lambda>2-\tilde{\eta}$ the interactions are
irrelevant with a scaling index $x=\tilde{\eta}$ (Fig.~\ref{fig:2}).
For $2<\lambda<2-\tilde{\eta}$ the interactions are still irrelevant
but now with the range dependent scaling index $x=2-\lambda$. In both
of these cases the fixed point structure is not changed and the
critical exponents take on their non-interacting values. However, even
these irrelevant interactions lead to a finite dissipative
conductivity at finite temperatures in contrast to the non-interacting
system \cite{Wea00}. For $\lambda<2$ and hence for Coulomb
interactions the interactions are relevant with scaling index
$x=2-\lambda$ and the system is driven away from the non-interacting
fixed point towards a new interacting fixed point \cite{LW96}.

\subsection{Scaling for Coulomb interactions}
\label{sec:Scal-Coul-inter}

Unfortunately, the linear stability analysis of the previous section
does not tell us what this new fixed point looks like, only that
infinitesimally weak interactions are sufficient to take us there. It
is known that for sufficiently strong interactions the QH system
develops new correlated ground states associated with the emergence of
the fractional QHE \cite{Lau83}. At these fixed points the interaction
strength is much larger than the disorder. These are not the fixed
points that we are concerned with here. We want to know the ground
state of the system for interaction strength much smaller than the
disorder strength.

In order to study the nature of the new fixed point a numerical
approach was chosen. Ideally, an exact method should be used but in
the presence of both interactions and disorder exact diagonalization
studies are limited to system sizes far too small to allow for any
finite-size scaling analysis to extract critical exponents. We
therefore treat the disorder exactly and incorporate the Coulomb
interactions in a self-consistent Hartree-Fock approximation (HFA).
The first calculation of this kind by Yang and MacDonald focused on
the tunneling density of states (DoS) and found a linear Coulomb gap
at all filling factors of the lowest Landau band, even at the critical
energy \cite{YM93}. In particular, it was found that the DoS
$\rho(E_F,L)$ at the Fermi energy in a system of linear dimension $L$
scales as
\begin{equation}
  \label{eq:2}
  \rho(E_F,L) \propto \frac{1}{\gamma L},
\end{equation}
where $\gamma=(e^2/4\pi\epsilon l)/\Gamma$ is a measure of the interaction
strength relative to the disorder strength $\Gamma$. Note that this
suppression of the DoS is basically of classical origin and not related 
to the quantum Hall critical point \cite{ES75}. 

This system size dependence of the DoS is in contrast to the size
independent DoS of the non-interacting system and offers an
explanation for a change of the dynamical critical exponent $z$ from 2
to 1. The dynamical exponent relates energy and length scales,
$E\propto L^{-z}$, and in a non-interacting system this relation is
given by the DoS. The corresponding energy and length scales are the
mean level spacing $\Delta$ and the system size $L$, respectively,
\begin{equation}
  \label{eq:3}
  \Delta = \frac{1}{\rho L^d},
\end{equation}
i.e. $z=d=2$. With interactions, however, the DoS is no longer constant
but scales like $1/L$ so that the dynamical exponent is reduced to
$z=d-1=1$. The suppression of the DoS is unrelated to the critical
point and happens for all values of the Fermi energy, but when the
Fermi energy coincides with the critical energy it changes the
critical exponent $z$.

It remains to be checked whether this simple argument holds and
whether other critical quantities are also changed by the interactions
or not. Using the self-consistent Hartree-Fock eigenvalues and
eigenfunctions, Yang, MacDonald, and Huckestein studied the scaling
behavior of the participation ratio and compared the Thouless numbers
of interacting and non-interacting systems \cite{YMH95}. The size and
energy dependence of the participation ratio could be fitted by the
same exponents $\nu$ and $D(2)$ that were obtained in the
non-interacting system. The Thouless numbers, that are in a
non-interacting system a measure of the conductance, also remained
unchanged upon inclusion of the interactions, although the relation
between the Thouless numbers and the conductance is unknown for an
interacting system.

\begin{figure}
  \hfill\psfig{figure=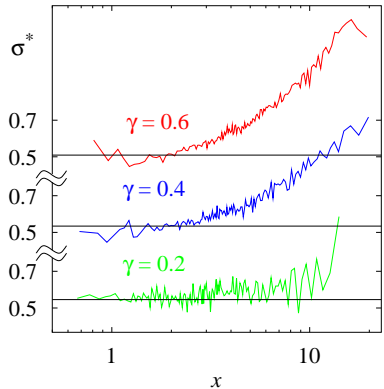,width=5cm}\hspace*{\fill}
  \caption[fig3]{Dynamical conductivity $\sigma^*$ in time-dependent
    Hartree-Fock approximation as a function of the scaling variable
    $x=q^2/\chi(q)\hbar\omega$ for different interactions strengths
    $\gamma$ (curves offset for clarity).}
  \label{fig:3}
\end{figure}

In order to directly obtain the conductivity and the dynamical scaling
one has to go beyond the self-consistent HFA and include vertex
corrections within the time-dependent Hartree-Fock approximation
(TDHF). This is the corresponding conserving approximation and while
the quantities calculated in the self-consistent HFA are
single-particle properties, the TDHF allows to calculate true
two-particle properties like the polarization and the
conductivity. A finite-size scaling analysis shows that in the limit
of vanishing frequency $\omega$ and wavevector $q$ the irreducible
polarization $\Pi(q,\omega)$ scales like
\begin{equation}
  \label{eq:4}
  \Pi(q,\omega) = \chi(q) \frac{\sigma^*(x) x }{\sigma^*(x) x -
    ie^2/\hbar}, 
\end{equation}
with $x=q^2/\chi(q)\hbar\omega$. The static susceptibility $\chi(q)$
vanishes like $q$ for small $q$, even when the vertex corrections are
included. Hence the scaling variable $x\propto q/\omega$ and the
dynamical critical exponent $z$ is indeed unity. Despite the vanishing
static susceptibility, the conductivity
$\sigma^*(x)=e^2\chi(q)D(q,\omega)$ is finite even in the DC limit as
the diffusion coefficient $D$ exhibits super-diffusive behavior and
diverges in this limit. Fig.~\ref{fig:3} shows the dynamical
conductivity $\sigma^*$ as a function of the scaling variable $x$
\cite{HB99}. The DC conductivity obtained in the limit $x\to0$ is
independent of the strength of the interaction and is given by
$\sigma_c=0.5\pm0.1e^2/h$. The scaling in the opposite limit
$x\to\infty$, $\sigma^*\propto x^{1-\eta}$, is governed by
multifractal correlation and allows to extract the anomalous diffusion
exponent $\eta=0.4\pm0.1$. Both of these values agree within the
uncertainties with their non-interacting counterparts.

The numerical calculations are consistent with the picture that the
interacting integer QH system is rather peculiar. The Coulomb
interaction is relevant at the non-interacting fixed point but drives
the system to a new fixed point, the Hartree-Fock fixed point, that
differs from the old one only by the occurrence of the Coulomb gap and
the associated reduction in the dynamical critical exponent $z$.
Whether this picture is correct or due to the failure of the
time-dependent Hartree-Fock approximation \cite{PS98} could be decided
by numerically calculating the scaling dimension of the residual
interactions at the HF fixed point, a rather daunting task that has
not been attacked so far. 

\section{Fate of the QHE at low magnetic fields}
\label{sec:Fate-QHE-at}

\begin{figure}
  \hfill\psfig{figure=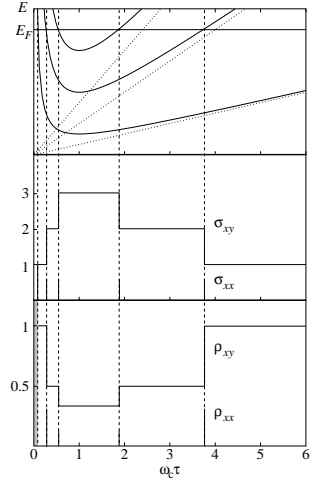,width=4cm}\hspace*{\fill}
  \caption[fig4]{Sketch of the magnetic field dependence of the critical
      energies according to the levitation picture
      ($E_n=(n+1/2)\hbar\omega_c(1+1/(\omega_c\tau)^2)$ \cite{Lau84})
      (top), the corresponding (dimensionless) conductivities
      (center), and resistivities (bottom).}
  \label{fig:4}
\end{figure}

The behavior of the critical energies, the conductivities, and the
resistivities as expected from the scaling theory of the IQHE is
sketched in Fig.~\ref{fig:4}. At high magnetic fields, the critical
energies are located close to the Landau energies. At each of the
critical points, the Hall conductivity changes by exactly 1 (from here
on, we measure all conductivities in units of $e^2/h$). As no extended
states exist below the Fermi energy at zero magnetic field, the
critical states can only move up in energy as the magnetic field goes
to zero \cite{Khm84,Lau84}. At fixed Fermi energy or particle number,
the sequence of Hall conductivities exhibited by
the system as the magnetic field is lowered is thus
0--1--2--\ldots--N--\ldots--2--1--0. In particular, the transition
from the last quantized plateau to the insulator at low fields can
only happen from the $n=1$ plateau.

It should be noted that the situation is fundamentally different in
lattice models of the QHE. While for weak disorder and strong magnetic
fields the low-lying magnetic subbands of a tight-binding band are
faithful approximations of the Landau bands of the continuum, lattices
models show completely different behavior at strong disorder/weak
magnetic fields due to the presence of states with negative Hall
conductivity that are absent in the continuum. These states can
combine with states carrying positive Hall conductivity and alter the
topology of the phase diagram in Fig.~\ref{fig:4}.

Since the dissipative conductivity is only finite at the plateau
transitions, the Hall resistivity exhibits as non-monotonic series of
plateaus, with the Hall resistivity increasing with decreasing
magnetic field below the $1/N$ plateau. This behavior predicted by
the scaling theory is in stark contrast to the experimental situation, 
where only monotonously increasing Hall resistivities are observed at
low magnetic fields. Even more disturbingly, a series of experiments
have observed a temperature independent point with insulating behavior
for lower magnetic fields and emerging quantized plateau with $n>1$
for higher magnetic fields
\cite{SKD93,KMFP95,Sea97,LCSL98,Hea99}. Obviously, if this temperature 
independent point signals the presence of a quantum critical point at
zero temperature, then the experiments are at odds with the phase
diagram of Fig.~\ref{fig:4}.

However, the observed linear, classical behavior of the Hall
resistivity at low fields gives us a hint how to properly interpret
the experimental data. First, we have to realize that the results of
scaling theory, such as the phase diagram of Fig.~\ref{fig:4}, deal
with the behavior of the system at zero temperature on asymptotically
large length scales. Experiments, on the other hand, are performed at
finite temperatures and a finite temperature introduces a finite
length scale $L_\Phi$, the phase coherence length. The question,
whether or not an experiment performed at a certain temperature
reflects the asymptotic scaling regime is thus a question of length
scales.

\begin{figure}
  \hfill\psfig{bbllx=0pt,bblly=388pt,bburx=431pt,bbury=884pt,figure=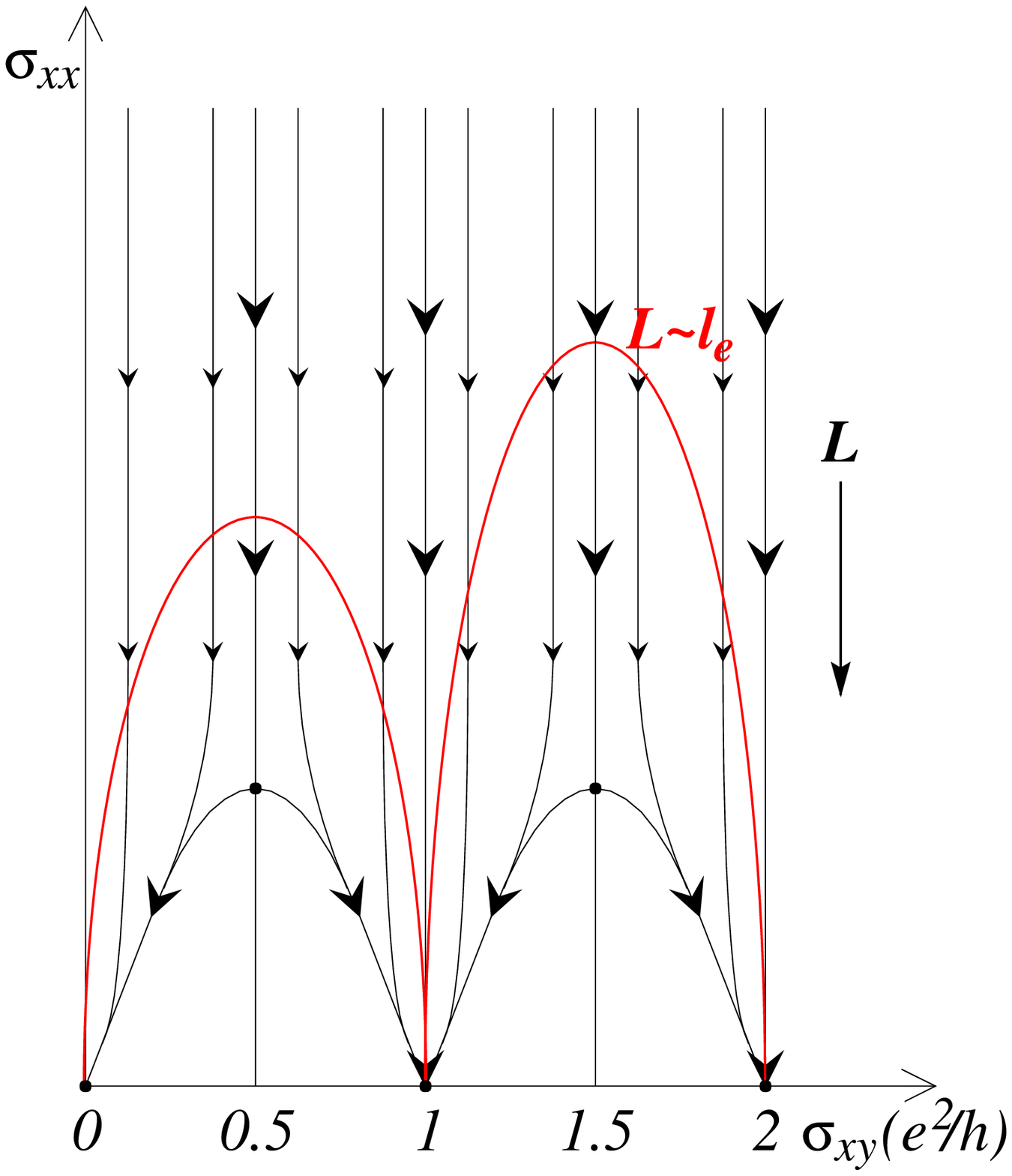,width=5cm,clip=}\hspace*{\fill}
  \caption[fig5]{Schematic flow diagram for the IQHE. The arrows
    indicate increasing length scale and the dashed curve represents
    the SCBA result.}
  \label{fig:5}
\end{figure}

How the conductivities change with length scale is determined by the
renormalization group $\beta$-functions of the system. The exact form
of these functions is not known for the QHE, only an approximation for
large longitudinal conductivity $\sigma_{xx}$ is given in the
literature \cite{Pru87}. The structure of the resulting flow diagram
is given in Fig.~\ref{fig:5} \cite{Khm83}. The flow starts at high
temperatures, corresponding to short length scales, with the
semiclassical values of the conductivity. The physics of the unstable
QH fixed points at half-integer $\sigma_{xy}$ becomes apparent at
length scales such that the longitudinal conductivity is of the order
of 1/2.

\begin{figure}
  \hfill\psfig{figure=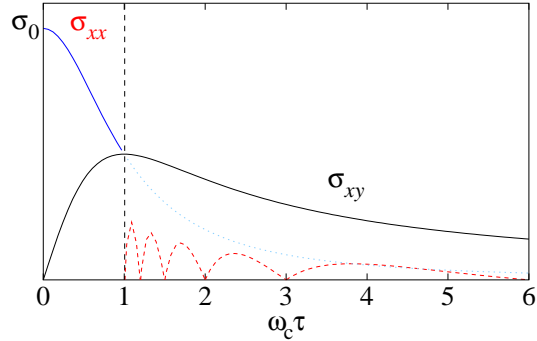,width=7cm}\hspace*{\fill}
  \caption[fig6]{Conductivities on short length scales: the Drude results
      (solid lines and dotted line) and the SCBA result for
      $\sigma_{xx}$ (dashed), appropriate for $\omega_c\tau\gg1$.}
  \label{fig:6}
\end{figure}

At high temperatures, the conductivities are given by the Drude
expressions
\begin{eqnarray}
  \sigma_{xx}^0 &=& \frac{\sigma_0}{1 + (\omega_c \tau)^2},\label{eq:10}\\
  \sigma_{xy}^0 &=& \omega_c\tau\sigma_{xx}^0,\label{eq:11}
\end{eqnarray}
with $\sigma_0=e^2 n_c \tau/m^*$, $\omega_c=eB/m^*$, and
$n_c$, $\tau=\ell/v_F$, and $\ell$ are the carrier density,
transport time, and the elastic mean free path, respectively
(Fig.~(\ref{fig:6})). The resulting linear increase of the Hall
resistivity is observed in all experiments. The change of the
longitudinal conductivity $\sigma_{xx}$ due to quantum interference
is given to lowest order in $1/\sigma_{xx}^0$ by the unitary
$\beta$-function \cite{Hik81,Efe83}
\begin{equation}
  \label{eq:uni_cor}
  \sigma_{xx}(L) = \sigma_{xx}^0 -
  \frac{1}{\pi^2\sigma_{xx}^0}\log
  \left(\frac{L}{l_c}\right).
\end{equation}

While this expression is only valid for large $\sigma_{xx}$, it can
provide a rough estimate of the length scale at which the weak
localization corrections become comparable to the bare value
$\sigma_{xx}^0$, at which point the influence of the QH fixed point
becomes dominant. Integrating eq.~(\ref{eq:uni_cor}), we get the
crossover length scale
\begin{equation}
  \label{eq:xi0}
  \xi^0=l_c \exp(\pi^2\sigma_{xx}^0{}^2).
\end{equation}
We note the strong dependence of the crossover scale on the bare
conductivity $\sigma{xx}^0$. In order to get a handle on the possible
values of $\sigma_{xx}^0$, we resort to Laughlin, who argued that the
critical energies are given by the condition, that the Drude
$\sigma_{xy}^0=(n+1/2)$ \cite{Lau84}, corresponding to the Landau
energies at high magnetic fields. In order to observe the $N$-th Hall
plateau, the maximum of $\sigma_{xy}^0$ at $\omega_c\tau=1$ needs to
be larger than $N-1/2$. Consequently, for magnetic fields up to
$\omega_c\tau=1$ the bare longitudinal conductivity $\sigma_{xx}^0$
exceeds $N-1/2$. In the experiments that claim to observe direct
transitions from higher plateaus, $N$ is at least 2. The corresponding 
crossover length scale $\xi^0$ is then at least several $10^9l_c$, a
macroscopic length, orders of magnitudes larger than any phase
coherence length achievable experimentally. We thus conclude, that at
low fields $\omega_c\tau<1$ the attainable length scales are too short 
to observe the QHE, even if it is present in an infinite
system. Instead, classical behavior with weak localization corrections 
is expected, in agreement with experiment.

So far, our discussion does not explain the experimentally observed
strong change in the temperature dependence of the resistivities near
$\omega_c\tau=1$. Instead of the weakly insulating temperature
dependence observed at low magnetic fields, at higher magnetic fields
the onset of plateau quantization is seen. In order to understand
this, we need to take into account the quantizing effect of the
magnetic field. For $\omega_c\tau>1$, the disorder broadening of the
Landau levels becomes smaller than their separation, leading to minima
in the density of states. These minima are accompanied by minima in
the bare conductivity $\sigma_{xx}^0$. For $\omega_c\tau\gg1$, the
Drude result in no longer valid and the bare conductivity is given by
the self-consistent Born approximation \cite{AU74}. At integer filling 
factors the bare conductivity becomes small and the crossover length
scale becomes microscopic, allowing the observation of QHE.

From these arguments, we expect an almost temperature independent
point near $\omega_c\tau=1$ separating a weakly temperature dependent, 
insulating regime at lower magnetic fields from a regime of stronger
temperature dependence with emerging quantized Hall plateaus, in
agreement with the experimental observations.

The present discussion leaves a lot of room for improvements. First,
we use only approximate expressions for the bare conductivities. Next,
the $\beta$-functions should be calculated all the way from large
conductivities to the QH fixed points. While improvements in both of
these areas are highly desirable, they are also hard to come by as the
combined effects of finite magnetic field, disorder and interactions
need to be taken into account. We should further distinguish between
transport and scattering times, but since the experiments are
performed on low mobility samples, there is not much difference. In
spite of the limitations, we believe that our discussion captures the
essential features of the physical situation.  Finally, we want to
point out that the whole discussion applies to samples that are
actually insulating at zero magnetic field. It certainly does not
apply to samples that are apparently metallic at zero field.

\section{Conclusions}
\label{sec:Conclusions}

The scaling theory of the plateau transitions in the integer quantum
Hall effect presents a picture that agrees with a wide range of
experimental findings. However, when analyzing experimental or
numerical data, it is imperative to consider the characteristic length 
scales in the system. This need becomes most prominent at low magnetic 
fields, where the emergence of an enormous crossover length scales
prohibits the experimental observation of the QHE.

Coulomb interactions appear to play a peculiar role in the IQHE.
Neglecting them, one obtains a theory that correctly reproduces the
observed plateaus and even most characteristic features of the
transitions. However, the Coulomb interaction is a relevant
perturbation at the non-interacting fixed point and the
non-interacting theory is not the correct scaling theory.
Incorporating the effects of the Coulomb interaction within a
self-consistent time-dependent Hartree-Fock approximation, we find that
only the dynamical critical exponent $z$ changes from 2 to 1, a change
that can be traced to the occurrence of a Coulomb gap in the tunneling
density of states.

\section{Acknowledgments}
\label{sec:Acknowledgements}

The support through the Sonderforschungsbereich 341 of the DFG is
gratefully acknowledged.


\end{document}